\documentclass[lettersize,journal]{IEEEtran}
\usepackage{amsmath, amsfonts, amsthm, amssymb}
\usepackage{enumitem}
\usepackage{algorithmic}
\usepackage{algorithm}
\usepackage{array}
\usepackage{stfloats}
\usepackage{textcomp}
\usepackage{stfloats}
\usepackage{stackengine}
\usepackage{url}
\usepackage{float} % 导入 float 宏包
\usepackage{verbatim}
\usepackage{graphicx}
\usepackage{threeparttable}
\usepackage{subcaption}
\usepackage{xcolor}

\usepackage[
    backend=biber,       % 使用biber后端
    style=ieee,       % 引用样式，可根据期刊要求更换 (e.g., ieee, apa, nature)
    citestyle=numeric-comp, % 将【正文中的引用】样式设为带压缩的数字
    maxcitenames=2,      % 正文引用时，超过2个作者就显示 "et al."
    maxbibnames=10,      % 参考文献列表中，超过10个作者才显示 "et al."
    minbibnames=10,   % 截断后，实际显示10个作者
    sorting=none         % 按引用顺序排序
]{biblatex}
\renewbibmacro*{url+urldate}{}
\AtEveryBibitem{%
  \clearfield{doi}%         移除 DOI
  \clearfield{isbn}%        移除 ISBN
  \clearfield{location}%    移除地点
  \clearfield{month}%       移除月份
  \clearfield{note}%        清除 "note" 字段，有时访问日期会被放在这里
}
\usepackage{hyperref}
\usepackage{booktabs}
\usepackage{supertabular} % <-- 核心
\usepackage{multirow}
\usepackage{afterpage}  % <-- 关键：引入 afterpage 宏包
\usepackage{cuted}
\usepackage{multirow} % Required for multi-row cells

\begin{document}

\title{Switching Efficiency: A Novel Framework for Dissecting AI Data Center Network Efficiency}

\author{
    Niangen Ye,
    Jiawen Zhu,
    Baojun Chen,
    Dong Wang,
    Jiang Sun,
    Weiqiang Sun, \IEEEmembership{Senior Member,~IEEE},
    and Weisheng Hu, \textit{Member,~IEEE}
    \thanks{N. Ye, J. Zhu, W. Sun, and W. Hu are with the State Key Laboratory of Photonics and Communications, Shanghai Jiao Tong University, Shanghai, China.}
    \thanks{B. Chen is with Pengcheng Laboratory, China.}
    \thanks{W. Dong and J. Sun are with Department of Fundamental Network Technology, China Mobile Research Institute, Beijing, China.}
    \thanks{Corresponding author: Weiqiang Sun (E-mail: sunwq@sjtu.edu.cn).}
}

% The paper headers
\markboth{Journal of \LaTeX\ Class Files,~Vol.~14, No.~8, August~2021}%
{Shell \MakeLowercase{\textit{et al.}}: A Sample Article Using IEEEtran.cls for IEEE Journals}

\maketitle

\begin{abstract}

Communication is pivotal in LLM training, and a thorough analysis of the communication efficiency of AI data center (AIDC) network is essential for guiding the design of these capital-intensive clusters. However, conventional metrics are inadequate for such analysis, as they do not directly link network activity to computational progress and lack granularity to diagnose the impact of different network design patterns. To address this, we introduce a metric framework, the Switching Efficiency Framework, whose core metric — Switching Efficiency ($\eta$) — quantifies computationally effective data throughput per unit switching capacity. We further decompose $\eta$ into three factors — Data, Routing Efficiency, and Port Utilization to facilitate analysis of distinct communication bottlenecks.

Using this metric framework, we demonstrate how the symmetric, distributed switching of 3D-Torus and the centralized, hierarchical switching of Rail-Optimized architecture align with sparse or imbalanced LLM training traffic, and show that All-to-All traffic from Mixture-of-Experts models severely degrades their port utilization and routing efficiency. Our analysis also demonstrates how key design choices — such as adjusting switching resource allocation, expanding server size, adopting in-network computing, and multi-plane design — positively influence distinct facets of communication efficiency. Ultimately, the Switching Efficiency Framework provides an analytical tool for analyzing efficiency bottlenecks, thereby informing the design of future-generation AIDC networks.

\end{abstract}

\begin{IEEEkeywords}
AI data center networks, Communication efficiency, Efficiency metric
\end{IEEEkeywords}

\section{Introduction}

The rapidly growing size of Large Language Models (LLMs) is driving the construction of hyperscale AIDCs, where tens or even hundreds of thousands of GPUs are interconnected by a high-speed network \cite{qian_alibaba_2024, jiang_megascale_2024, grattafiori_llama_2024, comanici_gemini_2025, meng_astral_2025, xai_colossus_}. Critically, with the growth in both model training volume and per-GPU computational capability outpacing the interconnect capabilities, communication has emerged as the principal performance bottleneck \cite{zhang_network_2020, erdil_data_2024, benyahya_mosaic_2025, wei_communication_2025}. This reality elevates the communication efficiency of the AIDC network to the central determinant of training throughput and system scalability for these multi-billion-dollar clusters \cite{cottier_rising_2025, pilz_trends_2025}.

To improve communication efficiency, existing studies have explored both algorithmic and architectural approaches. The former \cite{cowan_mscclang_2023, shah_taccl_2023, li_tccl_2024, sensi_swing_2024, song_innetwork_2024, zhao_adapcc_2024, zhao_forestcoll_2025} focuses on developing communication libraries and scheduling policies to minimize communication time during LLM training in existing network architectures. The latter \cite{dong_eflops_2020, hoefler_hammingmesh_2022, wang_topoopt_2023, jouppi_tpu_2023, nvidia_nvidia_2024b, qian_alibaba_2024, wang_railonly_2024, zu_resiliency_2024, liao_ubmesh_2025, yan_atop_2025, liao_mixnet_2025, shou_infinitehbd_2025} focuses on proposing domain-specific architectures to adapt to the structured nature (sparse and imbalanced) of the traffic in LLM training \cite{wang_railonly_2024, li_understanding_2024, liao_mixnet_2025}. As these approaches broaden the design space, it becomes increasingly critical to quantitatively assess how different design trade-offs impact the efficiency of AIDC networks, which is crucial in guiding the resource-efficient design of future million-GPU clusters.

\begin{figure}[h!]
    \centering
    \includegraphics[width=0.95\columnwidth]{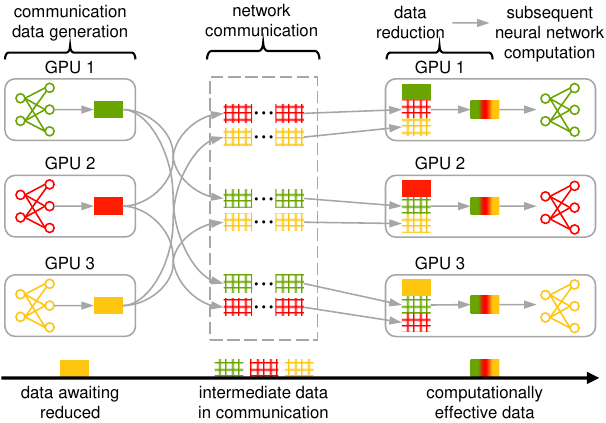} % Placeholder - use actual filename
    \caption{Intermediate data versus computationally effective data in implementing All-Reduce  without in-network computing.}
    \label{effective_data}
    \vspace{0pt}
\end{figure}

However, it is challenging to quantify the efficiency of AIDC networks in a way that disentangles design trade-offs to provide actionable guidance. First, in LLM training, network communication is a means to advance computation, not an end in itself. Consequently, the evaluation of communication efficiency should measure the network's direct contribution to computation. For instance, consider the All-Reduce operation in data parallelism of LLM training. As illustrated in Fig.~\ref{effective_data}, an implementation of All-Reduce without In-Network Computing (INC) would necessitate transmitting multiple intermediate data, which would be discarded after reduction to ``computationally effective data'' for subsequent neural network computation \cite{patarasuk_bandwidth_2009}. In contrast, an approach using INC delivers only the reduced result to the endpoints \cite{klenk_innetwork_2020} and thus exhibits higher efficiency. In this context, an efficiency evaluation that cannot distinguish between these two methods — that is, one that fails to measure the extent to which network resources are utilized to generate computationally effective data — provides an incomplete view.
\vspace{-5pt}
\vspace{-5pt}

\begin{figure}[h!]
    \centering
    \includegraphics[width=1\columnwidth]{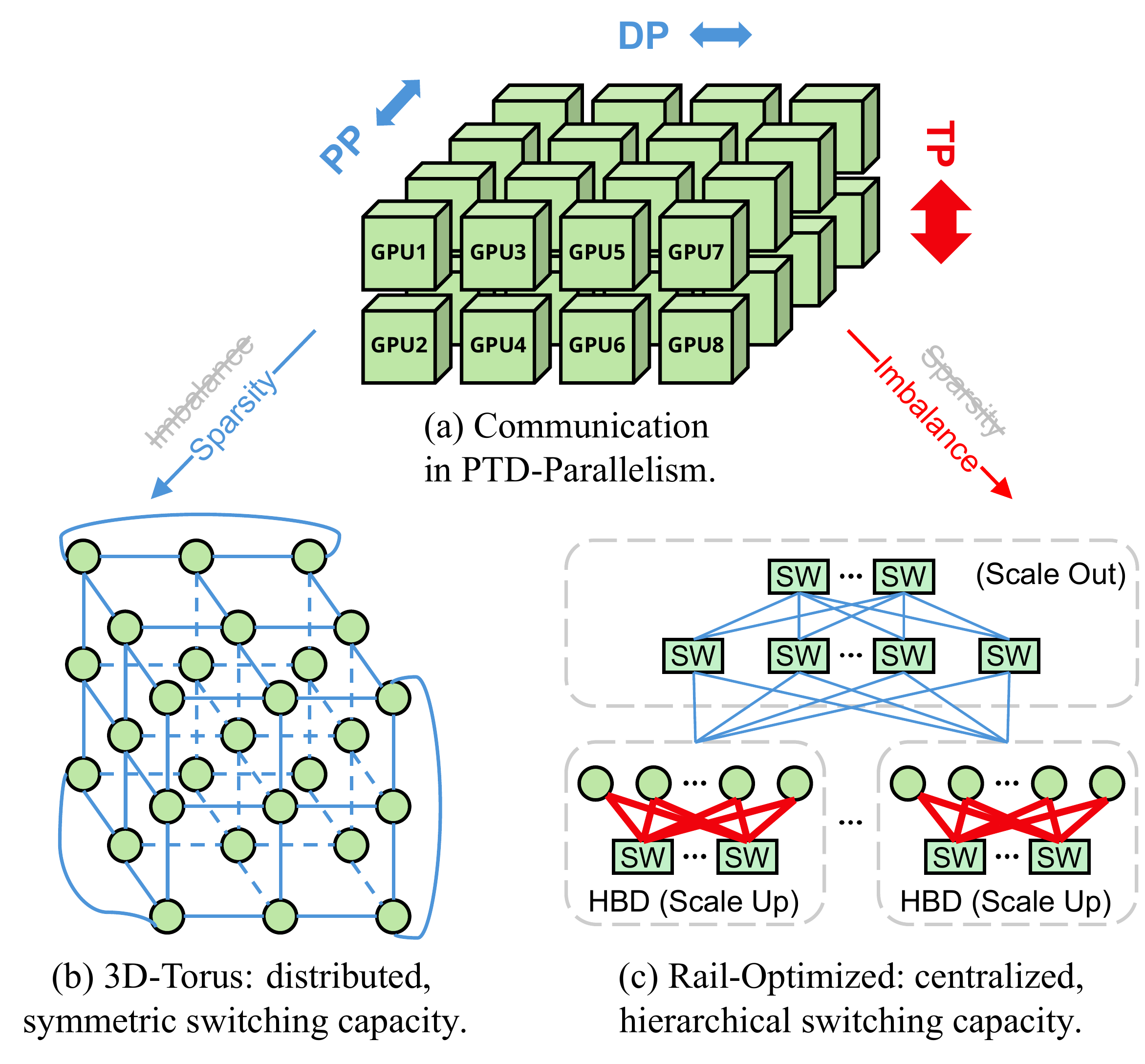} % Placeholder - use actual filename
    \caption{Selective alignment of topology with different LLM training traffic features in designing AIDC networks.}
    \label{traffic_driven_arch}
    \vspace{-3pt}
\end{figure}

Second, a prevailing design paradigm for AIDC networks is to tailor the topology to traffic patterns. An evaluation intended to guide network evolution should therefore quantify how different design trade-offs within that philosophy impact communication efficiency. Consider the 3D-Torus and Rail-Optimized architectures prevalent in AIDCs \cite{jouppi_tpu_2023, jiang_megascale_2024, xai_colossus_}. As illustrated in Fig.~\ref{traffic_driven_arch}, 3D-Torus matches the sparse communication patterns in PTD (Pipeline/Tensor/Data) parallelism by using low-radix switches (6 switching ports) within each accelerator, a design choice that avoids costly high-radix switches \cite{jouppi_tpu_2023, zu_resiliency_2024}. However, its symmetric, uniformly distributed bandwidth allocation mismatches the imbalance in LLM training traffic. Conversely, the Rail-Optimized architecture matches the imbalanced traffic by creating high-bandwidth ``scale-up'' domains (HBD) for intensive local communication (e.g., TP) and a ``scale-out'' fabric for lightweight remote communication (e.g., DP, PP) \cite{2022FastInterGPU, nvidia_scaling_}. Yet, its non-blocking connectivity design may be an over-provision for the sparse traffic. Crucially, while tailored for PTD parallelism, both design patterns are challenged by the all-to-all traffic in training Mixture-of-Experts (MoE) models \cite{comanici_gemini_2025, liu_deepseekv3_2024}. Given this background, an efficiency evaluation is ambiguous if it cannot separately quantify which efficiency dimensions of a design align with the specific traffic characteristics, as it would fail to provide clear guidance on how to evolve the architecture to resolve specific bottlenecks.
\vspace{-7pt}

Addressing these challenges requires developing an evaluation framework based on metrics precisely defined to capture the nature of the AIDC network. Here we argue that the design of this evaluation framework and its core metric should be based on the following principles:

\textit{\textbf{P1: Measure computationally effective data throughput.}} To link the communication efficiency of the AIDC network with its goal of advancing neural network computation, the core metric needs to quantify the throughput of computationally effective data that is immediately usable for subsequent neural network calculations. This application-level focus avoids merely tracking network ``busyness'' and instead assesses its direct contribution  to training progress.

\textit{\textbf {P2: Couple traffic with the configured switching capacity in network topology.}} To quantitatively guide the traffic-pattern-driven network design, the core metric needs to assess how effectively an architecture --- which physically manifests as a specific distribution of switching capacity in the topology --- is aligned with the structured traffic features of LLM training. By doing so, this metric enables comparison of key design trade-offs, such as different resource allocation strategies under different topologies from a resource utilization perspective, and thus promotes resource-efficient network design.

\textit{\textbf{P3: Enable diagnostic, fine-grained analysis.}} To further pinpoint how different design trade-offs impact communication efficiency in networks, the framework should also be able to dissect the macro-level efficiency into fine-grained factors. By enabling such fine-grained granularity, the evaluation framework turns these metrics from a simple efficiency benchmark to an analytical tool that reveals the sources of communication inefficiency, thereby providing better guidance for network design.

In this study, building on the above principles, we introduce the Switching Efficiency Framework, designed to analyze the communication efficiency of AIDC networks. The framework is centered on its core metric, Switching Efficiency ($\eta$), defined as the computationally effective data throughput of an LLM training workload to the total switching resources of the network. We then decompose $\eta$ into three fine-grained factors --- Data, Routing Efficiency, and Port Utilization ($\gamma,\delta,\theta$) --- each designed to isolate a specific bottleneck. Finally, we extend these metrics by introducing weighted efficiency that accounts for diverse LLM training workloads, enabling the analysis of long-term efficiency of production AIDCs that encounter various training workloads throughout their lifecycle.

Based on this framework, we analyze the communication efficiency of two network design patterns prevalent in hyper-scale AIDCs, represented by the 3D-Torus \cite{jouppi_tpu_2023, comanici_gemini_2025} and the Rail-Optimized architecture \cite{qian_alibaba_2024, jiang_megascale_2024, grattafiori_llama_2024, meng_astral_2025, xai_colossus_}. The analysis begins with the efficiency dissection across a diverse suite of dense and MoE model training workloads under fixed network architectures. We then extend the analysis to investigate how different network design options influence efficiency. Together, these analyses demonstrate the capacity of the Switching Efficiency Framework to both benchmark current architectures and inform the design direction of future AIDC networks. 

The major contributions of this work include the following:

\begin{itemize} [leftmargin=*, labelsep=5pt]
    \item We introduce a hierarchical metric framework, comprising a top-level metric ($\eta$) to quantify overall efficiency and a set of fine-grained factors ($\gamma,\delta,\theta$) to dissect the root inefficiency, providing actionable guidance for network design.
    \item We quantitatively reveal how the design patterns of 3D-Torus and Rail-Optimized effectively align with the sparse or imbalanced traffic of PTD-parallelism, yet are challenged by the all-to-all traffic in training MoE models.
    \item We clarify how different network design options, including switching resource allocation, server size, adoption of INC, and multi-plane design, impact the efficiency of port utilization, routing, and effective data reception.
\end{itemize}

%------------------- Literature Review -----------------%

\section{Literature Review}

\begin{table*}[!b]
\centering
\caption{Key Metrics for Analyzing Network Architectures Design in AIDC}
\label{tab:related_work_metrics}
\renewcommand{\arraystretch}{1.2}

\begin{tabular}{@{} p{2cm} p{5cm} p{3.1cm} p{2.0cm} @{}}
\toprule
\textbf{Category} & \textbf{Description} & \textbf{Typical Metrics} & \textbf{\hspace{0.0em}References} \\
\midrule

\multirow[t]{3}{*}{\parbox[t]{\linewidth}{\linespread{1.2}\selectfont \textbf{End-to-End\\Performance}}} &
\multirow[t]{3}{*}{\parbox[t]{\linewidth}{\linespread{1.2}\selectfont Quantifies the system-level performance from an application perspective.}} &
\parbox[t]{\linewidth}{\linespread{1.2}\selectfont Model FLOPs Utilization} &
\parbox[t]{\linewidth}{\linespread{1.2}\selectfont \raggedright\cite{jiang_megascale_2024, wang_railonly_2024, shou_infinitehbd_2025}} \\
\cmidrule(lr){3-4}
& &
\parbox[t]{\linewidth}{\linespread{1.2}\selectfont Sample Processing Rate} &
\parbox[t]{\linewidth}{\linespread{1.2}\selectfont \raggedright\cite{jiang_megascale_2024, khani_sipml_2021, jouppi_tpu_2023}} \\
\cmidrule(lr){3-4}
& &
\parbox[t]{\linewidth}{\linespread{1.2}\selectfont Iteration Time} &
\parbox[t]{\linewidth}{\linespread{1.2}\selectfont \raggedright\cite{jiang_megascale_2024, wang_railonly_2024, liao_mixnet_2025}} \\
\midrule

\multirow[t]{4}{*}{\parbox[t]{\linewidth}{\linespread{1.2}\selectfont \textbf{Communication\\Performance}}} &
\multirow[t]{4}{*}{\parbox[t]{\linewidth}{\linespread{1.2}\selectfont Evaluates the effectiveness and capacity of network fabric in supporting communication operations.}} &
\parbox[t]{\linewidth}{\linespread{1.2}\selectfont Collective Op. Bandwidth} &
\parbox[t]{\linewidth}{\linespread{1.2}\selectfont \raggedright\cite{hoefler_hammingmesh_2022, jouppi_tpu_2023, qian_alibaba_2024}} \\
\cmidrule(lr){3-4}
& &
\parbox[t]{\linewidth}{\linespread{1.2}\selectfont Network Throughput} &
\parbox[t]{\linewidth}{\linespread{1.2}\selectfont \raggedright\cite{jiang_unified_2020, liao_ubmesh_2025, qian_alibaba_2024}} \\
\cmidrule(lr){3-4}
& &
\parbox[t]{\linewidth}{\linespread{1.2}\selectfont Bandwidth Utilization} &
\parbox[t]{\linewidth}{\linespread{1.2}\selectfont \raggedright\cite{zhang_network_2020a, gangidi_rdma_2024, bonato_reps_2025}} \\
\cmidrule(lr){3-4}
& &
\parbox[t]{\linewidth}{\linespread{1.2}\selectfont Latency} &
\parbox[t]{\linewidth}{\linespread{1.2}\selectfont \raggedright\cite{dong_eflops_2020, khani_sipml_2021, gangidi_rdma_2024}} \\
\midrule

\multirow[t]{3}{*}{\parbox[t]{\linewidth}{\linespread{1.2}\selectfont \textbf{Topological\\Characteristic}}} &
\multirow[t]{3}{*}{\parbox[t]{\linewidth}{\linespread{1.2}\selectfont Describes the static, graph-theoretic properties of the physical layout of the network architecture.}} &
\parbox[t]{\linewidth}{\linespread{1.2}\selectfont Bisection Bandwidth} &
\parbox[t]{\linewidth}{\linespread{1.2}\selectfont \raggedright\cite{jouppi_tpu_2023, blach_highperformance_2024, feng_railx_2025}} \\
\cmidrule(lr){3-4}
& &
\parbox[t]{\linewidth}{\linespread{1.2}\selectfont Network Scale} &
\parbox[t]{\linewidth}{\linespread{1.2}\selectfont \raggedright\cite{qian_alibaba_2024, blach_highperformance_2024, han_lumoscore_2025}} \\
\cmidrule(lr){3-4}
& &
\parbox[t]{\linewidth}{\linespread{1.2}\selectfont Network Diameter} &
\parbox[t]{\linewidth}{\linespread{1.2}\selectfont \raggedright\cite{hoefler_hammingmesh_2022, feng_railx_2025}} \\
\cmidrule(lr){3-4}
& &
\parbox[t]{\linewidth}{\linespread{1.2}\selectfont Fault Explosion Radius} &
\parbox[t]{\linewidth}{\linespread{1.2}\selectfont \raggedright\cite{shou_infinitehbd_2025}} \\
\midrule

\multirow[t]{2}{*}{\parbox[t]{\linewidth}{\linespread{1.2}\selectfont \textbf{Capital\\Expenditure}}} &
\multirow[t]{2}{*}{\parbox[t]{\linewidth}{\linespread{1.2}\selectfont Measures the capital investment or energy footprint of a network architecture.}} &
\parbox[t]{\linewidth}{\linespread{1.2}\selectfont Cost \& Cost-Efficiency} &
\parbox[t]{\linewidth}{\linespread{1.2}\selectfont \raggedright\cite{hoefler_hammingmesh_2022, liu_lightwave_2023, liao_mixnet_2025}} \\
\cmidrule(lr){3-4}
& &
\parbox[t]{\linewidth}{\linespread{1.2}\selectfont Power \& Power-Efficiency} &
\parbox[t]{\linewidth}{\linespread{1.2}\selectfont \raggedright\cite{liu_lightwave_2023, wang_railonly_2024, shou_infinitehbd_2025}} \\
\bottomrule
\end{tabular}
\end{table*}

\subsection{Studies of Traffic Patterns and Communication Inefficiencies in AIDC Networks}

Many studies have empirically analyzed communication characteristics during LLM training to understand the sources of communication inefficiency in AIDC networks.

\textbf{Investigating the traffic patterns in LLM training.} Preliminary works investigated how different parallelism strategies influence the types and volumes of collective communication \cite{cheng_thorough_2024, anthony_demystifying_2024}. Further research delved deeper by analyzing the spatial distribution of traffic to characterize its properties at a more granular level \cite{li_understanding_2024, wang_railonly_2024, liao_mixnet_2025}. Specifically, the analyses in \cite{li_understanding_2024, wang_railonly_2024} exposed the sparse and imbalanced nature of LLM training traffic in PTD hybrid parallelism, which directly motivated the design of the sparsity-aware Rail-only topology \cite{wang_railonly_2024}. Similarly, the analysis in \cite{liao_mixnet_2025} revealed the locality of All-to-All traffic within Expert Parallelism (EP) for MoE models, leading to the proposal of MixNet, a hybrid optical-electrical architecture. While these studies provide foundational comprehension into LLM traffic characteristics, our work builds upon them by introducing a hierarchical metric framework to systematically analyze how different network design patterns align with these salient traffic patterns, aiming to derive actionable guidance for network design.

\textbf{Identifying communication inefficiencies in production.} \cite{qian_alibaba_2024, gangidi_rdma_2024, meng_astral_2025} point out that using ECMP routing policies can lead to hash collisions for LLM training traffic, particularly in the presence of link failures or when multiple tasks are training simultaneously. They exhibit that adopting a dual-plane Rail-Optimized architecture \cite{qian_alibaba_2024}, or employing traffic engineering and optimizing congestion control strategies \cite{meng_astral_2025} can mitigate hash collisions. ByteDance \cite{jiang_megascale_2024} identifies congestion control failure, link failure, and NCCL synchronization timeouts as common failure types that cause network inefficiency in its large-scale GPU clusters, and presents a tool to identify the causes of failures quickly. Collectively, these studies reveal a critical class of inefficiencies that arise from operational dynamics in production systems. Our work, on the other hand, introduces a metric framework to pinpoint the intrinsic efficiency of the network architecture itself, aiming at guiding the design of these capital-intensive clusters.

\subsection{Existing Metrics for Analyzing AIDC Network Design}

The analysis of AIDC network design is a multifaceted endeavor that draws upon a range of established metrics. As summarized in Table~\ref{tab:related_work_metrics}, existing metrics provide a basis for evaluating performance, topology, and cost. However, when it comes to analyzing how different design trade-offs impact the intrinsic efficiency of AIDC networks for LLM training, they exhibit the following limitations.

\textbf{Performance-oriented metrics offer limited diagnostic value for network design.} High-level, ``black-box" performance metrics such as collective operation bandwidth or iteration time reduce system performance to a monolithic score. While they can effectively report a symptom --- for instance, a performance bottleneck --- they cannot analyze its origin, be it an inefficient communication algorithm, a traffic-topology mismatch, or a resource allocation issue. Even a seemingly granular ``white-box" metric such as bandwidth utilization is deceptive because it fails to distinguish effective data transfers from wasteful traffic on congested multi-hop paths, consequently quantifying network busyness rather than useful throughput.

\textbf{Static topological metrics are traffic-agnostic, making them poor indicators of workload-specific performance.} These metrics, such as bisection bandwidth or network diameter, characterize the physical graph and are independent of the traffic it supports. However, this abstraction overlooks the highly structured and non-uniform traffic patterns intrinsic to LLM training. Consequently, such metrics cannot quantify the critical alignment between a topology and its intended workload. For example, a network may exhibit an acceptable high bisection bandwidth in the global domain, yet still create a severe bottleneck for intense, localized communication (e.g., traffic within a TP group) if its local connectivity is inefficient. These metrics, assessing only the static property, would fail to identify this traffic-specific bottleneck.

\textbf{Capital metrics, while crucial for deployment, are ill-suited for technical performance analysis.} These metrics are extrinsic economic attributes, rather than the intrinsic physical properties characterizing the system behavior. Consequently, they offer limited insight into resolving performance bottlenecks, which are rooted in physical rather than commercial limitations.

Collectively, these limitations underscore the need for a specific set of metrics designed for AIDC networks. In this article, we aim to address these limitations by proposing an evaluation framework with novel metrics, the details of which are provided in the subsequent sections.

\section{Design The Switching Efficiency Framework}

\begin{figure*}[!b]
    \centering
    \begin{subfigure}[c]{0.2\textwidth}
        \includegraphics[width=\linewidth]{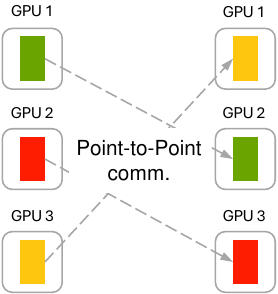}
        \caption{$\Delta_D = 3\cdot D$}
        \label{deltaD_0}
    \end{subfigure}  \hspace{0.2cm}
    \begin{subfigure}[c]{0.2\textwidth}
        \includegraphics[width=\linewidth]{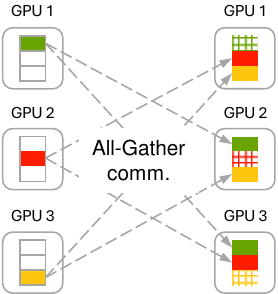}
        \caption{$\Delta_D = 3\cdot\frac{D}{3}\cdot(3 - 1)$}
        \label{deltaD_1}
    \end{subfigure}  \hspace{0.2cm}
    \begin{subfigure}[c]{0.2\textwidth}
        \includegraphics[width=\linewidth]{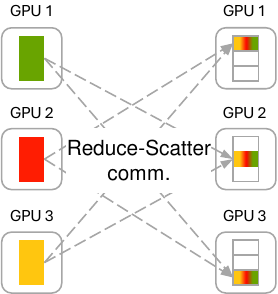}
        \caption{$\Delta_D = 3\cdot\frac{D}{3}$}
        \label{deltaD_2}
    \end{subfigure}\hspace{0.2cm}
    \begin{subfigure}[c]{0.2\textwidth}
        \includegraphics[width=\linewidth]{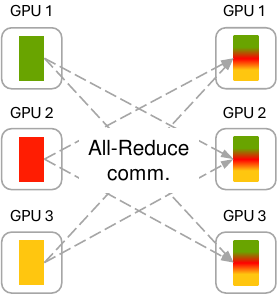}
        \caption{$\Delta_D = 3\cdot D$}
        \label{deltaD_3}
    \end{subfigure}\hspace{0.2cm}
    
    \begin{subfigure}[c]{0.41\textwidth}
        \includegraphics[width=\linewidth]{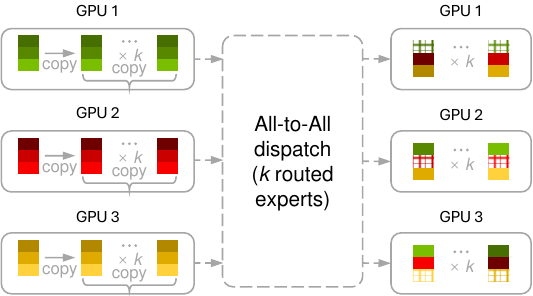}
        \caption{$\Delta_D = 3\cdot\frac{D}{3}\cdot(3 - 1)\cdot k_r$}
        \label{deltaD_4}
    \end{subfigure}\hspace{0.2cm}
    \begin{subfigure}[c]{0.41\textwidth}
        \includegraphics[width=\linewidth]{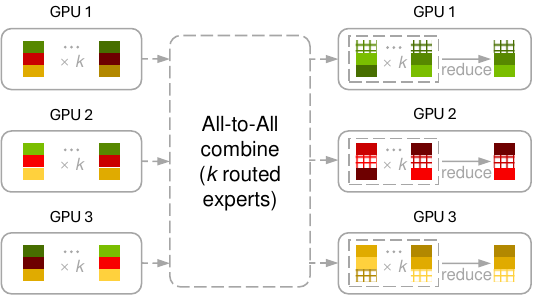}
        \caption{$\Delta_D = 3\cdot\frac{D}{3}\cdot(3 - 1)\cdot 1$}
        \label{deltaD_4}
    \end{subfigure}
    \caption{The effective data volume ($\Delta_D$) for each communication primitive. Assuming uniform token distribution for All-to-All dispatch and All-to-All combine. Grid-shaped shards represent the data that did not undergo network communication.}
    \label{deltaD}
    \vspace{-10pt}
\end{figure*}

Following the principles we have established, this section formally introduces the Switching Efficiency Framework, a metrics set for analyzing the communication efficiency of AIDC networks for LLM training.

\subsection{Designing the Core Metric: Switching Efficiency}

Guided by \textbf{P1} and \textbf{P2}, we define $\eta$ as the ratio of the total effective data (i.e., data that is directly usable for neural network computation after the completion of communication primitives, see Fig.~\ref{effective_data}) throughput to the network's aggregate switching capacity. Formally, let $\mathcal{P}$ be the set of all egress electrical switch ports, where each egress port $p \in \mathcal{P}$ has a data rate of $R_p$. For a network $\mathcal{A}$, consider a set of communication primitives $\mathcal{O}$ executed by the LLM training workload $w$ over a time duration $T$. If each primitive $i \in \mathcal{O}$ involves $N_i$ participants and yields an effective data volume of $\Delta_{D_i}$, the switching efficiency $\eta$ of network $\mathcal{A}$ for the workload $w$ is defined as:
\\
\begin{equation} \label{eq:eta}
\eta = \frac{\sum_{i\in \mathcal{O}} \Delta_{D_i}}{T \sum_{p\in {\mathcal{P}}} R_p}
\end{equation}
where
\\
\begin{equation} \label{eq:deltaD}
\Delta_{D_i} = 
\begin{cases} 
N_i \cdot D & \text{Point-to-Point} \\
N_i \cdot \frac{D}{N_i} \cdot (N_i - 1) & \text{All-Gather} \\ 
N_i \cdot \frac{D}{N_i} & \text{Reduce-Scatter} \\
N_i \cdot D & \text{All-Reduce} \\
N_i \cdot \frac{D}{N_i} \cdot (N_i - 1)\cdot k_r & \text{All-to-All dispatch} \\ 
N_i \cdot \frac{D}{N_i} \cdot (N_i - 1)\cdot 1 & \text{All-to-All combine} \\ 
\end{cases}
\end{equation}
\\
in which $\mathbf{\textit{D}}$ denotes the size of the full gradient or activation tensor on each GPU that the communication primitive acts on, and $k_r$ denotes the number of routed experts. And for a two-layer Rail-Optimized architecture,
\\
\begin{equation} \label{eq:switching_Resource}
\sum_{p\in \mathcal{P}} R_p = \sum_{p \in \mathcal{R}_S} R_p + \sum_{p \in \mathcal{R}_L} R_p + \sum_{p \in \mathcal{R}_{SV}} R_p + \sum_{p \in \mathcal{R}_G} R_p
\end{equation}
\\
in which  $\mathcal{R}_S$, $\mathcal{R}_L$, $\mathcal{R}_{SV}$, and $\mathcal{R}_G$ denote the sets of switch ports at Spine and Leaf layers, and within servers and GPUs, respectively.

Here, the effective data volume $\Delta_{D_i}$ is defined as the increment of data that is directly usable to continue computations of the neural network, as illustrated in Fig.~\ref{deltaD}. Specifically, the exact calculation of $\Delta_{D_i}$ under different primitives can be categorized as:

\textbf{1.~Non-reduction primitives} (Point-to-Point, All-Gather, All-to-All dispatch): All received data is directly usable for subsequent neural network computation. Therefore, $\Delta_{D_i}$ equals the aggregate newly received data volume. For instance, in All-to-All dispatch, $N_i$ GPU receives $N_i\cdot\frac{D}{N_i} \cdot (N_i - 1) \cdot k_r$ unique shards, yielding $\Delta_{D_i} = N_i \cdot \frac{D}{N_i} \cdot (N_i - 1) \cdot k_r$.

\textbf{2.~Reduction primitives} (Reduce-Scatter, All-Reduce, All-to-All combine): Data is usable only after reduction. Therefore, $\Delta_{D_i}$ is calculated based solely on the final reduced output. For instance, in All-to-All combine, although its raw communication volume might match that of All-to-All dispatch (without INC), each GPU ultimately retains only its final reduced shards, giving $\Delta_{D_i} = N_i \cdot \frac{D}{N_i} \cdot (N_i - 1) \cdot 1$.

\begin{figure}[h!]
    \centering
    \includegraphics[width=1\columnwidth]{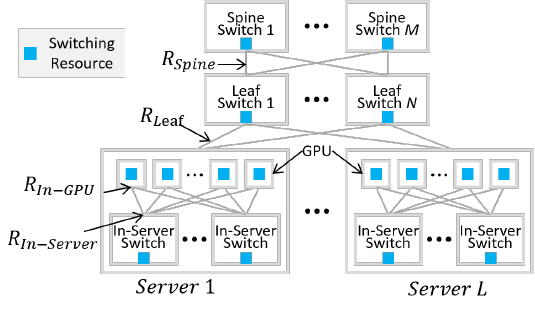} % Placeholder - use actual filename
    \caption{In a two-layer AIDC network, switching resources are
distributed in switches, on server boards, and in GPUs.}
    \label{fig:switching_resource}
    \vspace{-5pt}
\end{figure}

On the other hand, $\sum_p R_p$ is defined as the aggregate data rate of all electrical switch ports in the network $\mathcal{A}$, including those within GPUs capable of data forwarding, as exemplified by the two-layer Rail-Optimized architecture in Fig.~\ref{fig:switching_resource}.

In essence, Switching Efficiency ($\eta$) directly measures how effectively a network architecture can utilize its total provisioned switching capacity to transfer computationally useful data. By relating the workload-specific effective data throughput ($\sum \Delta_{D_i}$) to the architecture-defined aggregate switching resource ($\sum R_p$), this metric serves as a top-level indicator of the fundamental alignment between the workload and the network design.
% \vspace{-5pt}
\subsection{Decomposing Switching Efficiency for Fine-Grained Bottleneck Analysis}

Although $\eta$ provides a holistic score, it encompasses the effects of communication operation, network topology, and resource configuration. To enable a more granular, diagnostic capability (\textbf{P3}), we decompose $\eta$ to facilitate fine-grained bottleneck analysis by isolating and quantifying the efficiency of different contributing factors.

Formally, let $D^{\text{recv}}_g$ be the data volume received by GPU $g$ (Let $\mathcal{G}$ denote the set of GPUs) during time $T$, and $f_p(t)$ be the forwarding rate of egress port $p$ at time $t$. $\eta$ is decomposed as follows:
\\
\begin{equation} \label{eq:eta_decompose}
\begin{aligned}
\eta &= \underbrace{\frac{\sum_{i \in \mathcal{O}} \Delta_{D_i}}{\sum_{g \in \mathcal{G}} D^{\text{recv}}_{g}}}_{\gamma:\text{Data Efficiency}} \cdot \underbrace{\frac{\sum_{g\in \mathcal{G}} D^{\text{recv}}_{g}}{\sum_{p\in \mathcal{P}} R_p T}}_{\mu:\text{Network Efficiency}}
\\
& = \underbrace{\frac{\sum_{i} \Delta_{D_i}}{\sum_{g} D^{\text{recv}}_{g}}}_{\gamma:\text{Data Efficiency}} \cdot \underbrace{\frac{\sum_{g} D^{\text{recv}}_{g}}{\sum_{p} \int_{0}^{T} f_p(t)dt}}_{\delta:\text{Routing Efficiency}} \cdot \underbrace{\frac{\sum_{p} \int_{0}^{T} f_p(t)dt}{\sum_{p} R_p T}}_{\theta:\text{Port Utilization}}
\end{aligned}
\end{equation}
\\
where the composite metric $\mu$ (\textbf{Network Efficiency}) provides a network-level perspective on efficiency, and the fine-grained factors pinpoint the specific type of efficiency bottleneck, as elaborated below: % 先说一下每个指标的定义，作用，目的（宏观）；进一步再说这个东西实际上是一个什么东西；

\textbf{1.~Data Efficiency ($\boldsymbol{\gamma}$)}: This factor is defined as the ratio of the total effective data increment ($\sum \Delta_{D_i}$) to the total data volume received by all GPUs ($\sum D^{\text{recv}}_{g}$). It highlights the inefficiency stemming from the redundancy in the implementation of communication. Essentially, it reveals data redundancy ($r_\text{byte}$) caused by redundant reception within the implementation of communication primitives:
\\
\begin{equation} \label{eq:gamma_model}
\begin{aligned}
\gamma &= \frac{\sum_{i\in \mathcal{O}} \Delta_{D_i}}{\sum_{g\in \mathcal{G}} D^{\text{recv}}_{g}} \\&= \frac{1}{1+\frac{\sum_{g} D^{\text{recv}}_{g}-\sum_{i} \Delta_{D_i}}{\sum_{i} \Delta_{D_i}}} = \frac{1}{1+r_\text{byte}}
\end{aligned}
\end{equation}
\\
where $r_\text{byte}$ represents the redundant bytes that must be received to yield one byte of effective data. A low $\gamma$ indicates inefficient implementation of communication primitives, exemplified by operations like All-to-All combine and All-Reduce without INC, where a high $r_\text{byte}$ arises as GPUs receive extra intermediate data that is discarded after reduction.

\textbf{2.~Routing Efficiency ($\boldsymbol{\delta}$)}: This factor is defined as the ratio of the total data volume received by all GPUs to the aggregate data volume forwarded by all switch ports ($\sum \int_{0}^{T} f_p(t)dt$). It indicates how well the traffic flows can be routed in the network topology. Equivalently, $\delta$ is the inverse of the volume-weighted average number of switch-forwarding actions $\bar{n}_{\text{fwd}}$ of the data flows
 (assuming that the total data sent equals the total data received for all GPUs, and that each switch satisfies flow conservation):
\\
\begin{equation} \label{eq:delta_model}
\begin{aligned}
\delta
&= \frac{\sum_{g\in \mathcal{G}} D^{\text{recv}}_{g}}{\sum_{p\in \mathcal{P}} \int_{0}^{T} f_p(t)dt}\\
&= \frac{\sum_{f \in \mathcal{F}} V_f}{\sum_{f \in \mathcal{F}} (V_f \cdot n_f)}
= \frac{1}{\frac{\sum_{f \in \mathcal{F}} (V_f \cdot n_f)}{\sum_{f \in \mathcal{F}} V_f}}
= \frac{1}{\bar{n}_{\text{fwd}}}
\end{aligned}
\end{equation}
\\
where $\mathcal{F}$ is the set of all data flows, $V_f$ is the volume of a flow $f$, and $n_f$ is its number of switch-forwarding actions. The term $\bar{n}_{\text{fwd}}$ is thus the volume-weighted switch-forwarding count. A low $\delta$ value, therefore, directly reveals the case of topological mismatch that forces significant data volume to undergo multiple forwarding actions, a situation corresponding to a high $\bar{n}_{\text{fwd}}$. 

\textbf{3.~Port Utilization ($\boldsymbol{\theta}$)}: This factor is defined as the ratio of the aggregate data forwarding rate on all switch ports to the aggregate switching capacity time product ($\sum R_pT$). It assesses the utilization efficiency of provisioned resources on the network topology by revealing efficiency across both the spatial ($\theta_{\text{spatial}}$) and temporal ($\theta_{\text{temporal}}$) dimensions 
(Letting $\mathcal{P}_{\text{active}} \subseteq \mathcal{P}$ be the set of active ports with non-zero forwarding rates, i.e., $f_p(t) > 0$):
\begin{equation} \label{eq:theta_model}
\resizebox{0.44\textwidth}{!}{$
\begin{aligned}
\theta &= \frac{\sum_{p\in \mathcal{P}} \int_{0}^{T} f_p(t)dt}{\sum_{p\in \mathcal{P}} R_p T}\\
&= \frac{\sum_{p \in \mathcal{P}_{\text{active}}} R_p}{\sum_{p \in \mathcal{P}} R_p} \cdot  \frac{\sum_{p \in \mathcal{P}_{\text{active}}}\int_{0}^{T}  f_p(t)dt}{T \sum_{p \in \mathcal{P}_{\text{active}}} R_p} = \theta_{\text{spatial}}\cdot \theta_{\text{temporal}}
\end{aligned}
$}
\end{equation}
\\
where $\theta_{\text{spatial}}$ represents the fraction of total deployed port capacity that is activated, and $\theta_{\text{temporal}}$ represents the intensity of that activated capacity utilized for data forwarding. A low $\theta$ value thus indicates inefficient spatial or temporal resource configuration. For instance, an inappropriate allocation of dedicated resources for different communication phases (e.g., providing insufficient bandwidth for TP but excessive for PP) would degrade the utilization of alternately activated network subsystems, dually manifests as a reduced $\theta_{\text{spatial}}$ within each phase and a lowered $\theta_{\text{temporal}}$ across the entire iteration.

Table~\ref{tab:bottleneck_factors_summary} demonstrates the application of the fine-grained metrics in analyzing principal efficiency bottlenecks within typical scenarios, highlighting how each metric reveals specific inefficiencies arising from the implementation of communication primitives, network topologies, or resource allocations.

\vspace{-3.1pt}

\begin{table}[h!]
\begin{threeparttable}
\centering
\caption{Bottleneck examples in AIDC identified by the fine-grained metrics}
\label{tab:bottleneck_factors_summary}
\renewcommand{\arraystretch}{1.2}

\begin{tabular}{@{} p{0.19\columnwidth} p{0.76\columnwidth} @{}}
\toprule

\parbox[t]{\linewidth}{\linespread{1.2}\selectfont \textbf{Metrics}} &
% --- Markers are now placed in the column header here ---
\parbox[t]{\linewidth}{\linespread{1.2}\selectfont \textbf{Examples of Identified Bottleneck}} \\
\midrule

\parbox[t]{\linewidth}{\linespread{1.5}\selectfont \textbf{Data \\ Efficiency ($\gamma$)}
} &
% --- Markers removed from the table body ---
\parbox[t]{\linewidth}{\linespread{1.5}\selectfont \textbf{Redundant reception in reductions without INC:\tnote{\textdagger}}
\newline
\textbullet \hspace{0.2em} Reduce-Scatter (Recv: $n-1$ byte, Eff: $1$ byte):\\
\phantom{\textbullet \hspace{0.2em}} 
$r_\text{byte}=n-2 \to \gamma=\frac{1}{n-1}$
\newline
\textbullet \hspace{0.2em} All-Reduce (Recv: $2(n-1)$ byte, Eff: $n$ byte):\\
\phantom{\textbullet \hspace{0.2em}} 
$r_\text{byte}=\frac{n-2}{n} \to \gamma=\frac{n}{2(n-1)}$
\newline
\textbullet \hspace{0.2em} A2A combine (Recv: $(n-1)k_r$ byte, Eff: $n-1$  byte): \\ \phantom{\textbullet \hspace{0.2em}}
$r_\text{byte}=k_r-1 \to \gamma=\frac{1}{k_r}$}\\
\midrule

\parbox[t]{\linewidth}{\linespread{1.5}\selectfont \textbf{Routing \\ Efficiency ($\delta$)}
} &
\parbox[t]{\linewidth}{\linespread{1.3}\selectfont 
\textbf{Traffic requires multi-forwardings within the network:}

\textbullet \hspace{0.2em} Inter-Pod traffic over Leaf-Spine-Core-Spine-Leaf:\\ \phantom{\textbullet \hspace{0.2em}} $\bar{n}_{\text{fwd}} \approx 5 \to \delta \approx \frac{1}{5}$
\newline
\textbullet \hspace{0.2em} Inter-server All-to-All traffic over Leaf-Spine-Leaf:\\ \phantom{\textbullet \hspace{0.2em}} $\bar{n}_{\text{fwd}} \approx 3 \to \delta \approx \frac{1}{3}$
\newline
\textbullet \hspace{0.2em} All-to-All traffic along one dimension on 3D-Torus:\\
\phantom{\textbullet \hspace{0.2em}} 
$\bar{n}_{\text{fwd}} \propto k_r \to \delta \propto \frac{1}{k_r}$
} \\
\midrule

\parbox[t]{\linewidth}{\linespread{1.5}\selectfont \textbf{Port \\ Utilization ($\theta$)}
} &
\parbox[t]{\linewidth}{\linespread{1.3}\selectfont
\textbf{Allocation of one-dimension switching port resources on a 3D-Torus to TP/EP traffic in 3D parallelism:}

\textbullet \hspace{0.2em} TP phase (utilizes bidirectional link via ring algorithm): \\
\phantom{\textbullet \hspace{0.2em}} $\theta_{\text{spatial}}= \frac{2}{6}, \theta_{\text{temporal}} \approx 1 \to \theta \approx \frac{1}{3}$

\textbullet \hspace{0.2em} EP phase (utilizes unidirectional link w/o ring algorithm): \\
\phantom{\textbullet \hspace{0.2em}} $\theta_{\text{spatial}} = \frac{1}{6}, \theta_{\text{temporal}} \approx 1 \to \theta \approx \frac{1}{6}$
} \\
\bottomrule
\end{tabular}

\begin{tablenotes}[flushleft]
    \item[\textdagger] $n$: number of communication participants; $k_r$: number of routed experts. The received bytes and the effective bytes are counted across $n$ GPUs; see~\cite{li_understanding_2024, jin_megascalemoe_2025} and Eq.~\eqref{eq:deltaD} for the detailed formulas for each primitive.
\end{tablenotes}
\end{threeparttable}
\end{table}

\subsection{Generalizing Efficiency Metrics for Long-term Analysis}

To further assess production AIDCs that typically encounter heterogeneous LLM training workloads with distinct traffic patterns in their life cycle, we generalize the definition of the efficiency metrics for a single workload to a workload profile. Formally, consider a set of $K$ distinct training workloads, denoted as $\mathcal{W} = \{w_1, w_2, \dots, w_K\}$ that occur within an observation window $T_H$, the generalized form of Switching Efficiency $\eta$ is defined as:
\\
\begin{equation}   \label{eq:eta_generalized}
\begin{aligned}
\bar{\eta} = \frac{\sum_{k=1}^{K} \sum_{i \in \mathcal{O}_{w_k}} \Delta_{D_{i, w_k}}}{T_H \sum_{p \in \mathcal{P}} R_p}
\end{aligned}
\end{equation}
\\
Correspondingly, the generalized form of the efficiency metrics $\bar{\gamma}$, $\bar{\delta}$, $\bar{\theta}$, and $\bar{\mu}$ can be derived analogously by applying the decomposition logic in Eq.~\eqref{eq:eta_decompose}.
 
Specifically, assuming that workloads occur sequentially without overlap, such that $T_H = \sum_{k=1}^K T_k=T_\text{total}$ where $T_k$ is the duration of workload $w_k$, Eq.~\eqref{eq:eta_generalized} can be reformulated into a more concise and intuitive form:
\\
\begin{equation} \label{eq:eta_generalized_derivation_concise}
\begin{aligned}
\bar{\eta} &= \frac{\sum_{k=1}^{K} \sum_{i \in \mathcal{O}_{w_k}} \Delta_{D_{i, w_k}}}{T_H \sum_{p \in \mathcal{P}} R_p} \\
&= \sum_{k=1}^{K} \frac{T_k}{T_H} \left( \frac{\sum_{i \in \mathcal{O}_{w_k}} \Delta_{D_{i, w_k}}}{T_k \sum_{p \in \mathcal{P}} R_p} \right) = \sum_{k=1}^{K} \frac{T_k}{T_\text{total}} \cdot \eta_k\\ &= \sum_{k=1}^{K} \lambda_k \cdot \eta_k
\end{aligned}
\end{equation}
\\
This simplified form decouples the analysis from a specific observation window $T_H$, allowing the overall efficiency to be computed as a weighted sum of individual efficiencies $\eta_k$ using the time-based weights $\lambda_k$. This facilitates the theoretical analysis and comparison of network architectures under standardized or projected operational conditions within the life cycle of the AIDCs. Nevertheless, in scenarios with concurrent workloads or resource contention, direct measurement via Eq.~\eqref{eq:eta_generalized} is required to maintain accuracy.

\subsection{The Switching Efficiency Framework: A Hierarchical Metric Set for AIDC Network Analysis}

Collectively, the metrics defined in this section constitute the \textit{Switching Efficiency Framework}, providing a hierarchical structure for top-down network communication efficiency analysis. At the holistic level, $\eta$ and $\mu$ offer a top-level assessment of communication effectiveness. The fine-grained metrics ($\gamma, \delta, \theta$) then facilitate a diagnostic analysis, isolating bottlenecks related to the communication algorithm, network topology, and resource allocation. Finally, the generalized metrics ($\bar{\eta}$, $\bar{\mu}$, $\bar{\gamma}$, $\bar{\delta}, \bar{\theta}$) extend these analytical capabilities to long-term efficiency assessments under diverse production workloads. Table~\ref{tab:switching_efficiency_framework} summarizes the components of this framework and their analysis focus.

\begin{table}[ht]
\centering
\caption{Components in Switching Efficiency Framework}
\label{tab:switching_efficiency_framework}
\renewcommand{\arraystretch}{1.2}

\begin{tabular}{@{} p{0.18\columnwidth} p{0.11\columnwidth} p{0.61\columnwidth} @{}}
\toprule
\textbf{Level} & \textbf{Metrics} & \textbf{Analysis Focus} \\
\midrule

\multirow[t]{2}{*}{\parbox[t]{\linewidth}{\linespread{1.2}\selectfont \textbf{Holistic}}} &
\parbox[t]{\linewidth}{\linespread{1.2}\selectfont $\eta$} &
\parbox[t]{\linewidth}{\linespread{1.2}\selectfont Effectiveness of end-to-end communication} \\
\cmidrule(lr){2-3}
&
\parbox[t]{\linewidth}{\linespread{1.2}\selectfont $\mu$} &
\parbox[t]{\linewidth}{\linespread{1.2}\selectfont Effectiveness of network-level communication} \\

\midrule

\multirow[t]{3}{*}{\parbox[t]{\linewidth}{\linespread{1.2}\selectfont \textbf{Fine-grained}}} &
\parbox[t]{\linewidth}{\linespread{1.2}\selectfont $\gamma$} &
\parbox[t]{\linewidth}{\linespread{1.2}\selectfont Data redundancy in communication algorithm} \\
\cmidrule(lr){2-3}
&
\parbox[t]{\linewidth}{\linespread{1.2}\selectfont $\delta$} &
\parbox[t]{\linewidth}{\linespread{1.2}\selectfont Data forwarding count imposed by topology} \\
\cmidrule(lr){2-3}
&
\parbox[t]{\linewidth}{\linespread{1.2}\selectfont $\theta$} &
\parbox[t]{\linewidth}{\linespread{1.2}\selectfont Utilization of switching resource} \\

\midrule

\multirow[t]{1}{*}{\parbox[t]{\linewidth}{\linespread{1.2}\selectfont \textbf{Long-term}}} &
\parbox[t]{\linewidth}{\linespread{1.2}\selectfont $\bar{\eta}$, $\bar{\mu}$, $\bar{\gamma}$, $\bar{\delta}$, $\bar{\theta}$} &
\parbox[t]{\linewidth}{\linespread{1.2}\selectfont Long-term efficiency under diverse workloads} \\

\bottomrule
\end{tabular}
\end{table}

\section{Analyzing Communication Efficiency with the Switching Efficiency Framework under LLM training workloads}

Having established the theoretical foundation of the switching efficiency framework, this section demonstrates its practical utility. Based on the simulation setup detailed below, this section applies the proposed switching efficiency metric framework to dissect the communication efficiency of the production network architectures under various LLM workloads and network design parameters.

\subsection{Simulation Setup}

\textbf{1. Workload Specification.} A suite of workloads is specified based on hybrid parallelism strategies for both dense and MoE models. Specifically, for dense models, we employ a 3D (DP, PP, TP) strategy, with their parallelism degrees denoted by the tuple $(d, p, t)$. For MoE models, we adopt an approach similar to that of the training of DeepSeek-V3, which replaces TP with EP and decouples DP between Attention and MoE layers, resulting in two configurations: a 2D (DP, PP) strategy for Attention layers with degrees ($d,p$), and a 3D (EDP, PP, EP) strategy for MoE layers with degrees ($d_e,p,e$). Based on these strategies, a set of workload configurations is then specified through the following procedure:

\textit{Step 1: Enumerate Parallelism Configurations.} All valid parallel configurations for a cluster of size $N$ (total GPUs) are first enumerated. For dense models, a valid configuration is represented by a tuple $(d, p, t)$ such that $d \cdot p \cdot t = N$. For MoE models, initial tuples $(d_e, p, e)$ for the MoE layer are first identified such that $d_e\cdot p\cdot e = N$. The Attention layer's DP degree $d$ is then determined by the relationship $d = d_e\cdot e$, ensuring $d\cdot p = N$.

\textit{Step 2: Apply Practical Constraints.} To ensure realism, several constraints derived from established training practice are imposed on these configurations. The size of PP is constrained between 2 (to avoid overly shallow models) and $\frac{N}{16}$ (to prevent excessively deep pipelines). Similarly, the size of EP is constrained between 2 (with $\frac{e}{2}$ routed experts to ensure the presence of All-to-All communication) and $\frac{N}{32}$ (to avoid the formation of impractically large EP groups). Additionally, following typical hardware placement constraints, the maximum TP size is limited to the number of GPUs within a single high-bandwidth domain --- i.e., one server in the Rail-Optimized architecture (e.g., 8 GPUs).

\textit{Step 3: Scale Workload Parameters.} For each parallelism configuration, a specific LLM workload is specified by scaling model architectural parameters. The number of layers, hidden dimension size, number of experts (for MoE model), and batch size are set to be respectively proportional to the parallelism degrees $p, t, e,$ and the DP degree of attention layer $d$. The scaling coefficients are derived from the two typical model architectures: GPT-3 (for dense model) and DeepSeek-V3 (for MoE model), ensuring that the specified workloads serve as scaled analogues of these seminal architectures. The overall procedure to specify the workloads is summarized in Table~\ref{tab:workload_specification_summary}.

\begin{table}[h!]
\centering
\caption{Summary of Workload Specification}
\label{tab:workload_specification_summary}
\renewcommand{\arraystretch}{1.2} %
% \normalsize 

\begin{tabular}{@{}p{0.18\columnwidth} p{0.77 \columnwidth}@{}}
\toprule

\parbox[t]{\linewidth}{\linespread{1.2}\selectfont \textbf{Parameter}} & 
\parbox[t]{\linewidth}{\linespread{1.2}\selectfont \textbf{Configuration Setting}} \\
\midrule

\parbox[t]{\linewidth}{\linespread{1.2}\selectfont \textbf{Model Types}} & 
\parbox[t]{\linewidth}{\linespread{1.2}\selectfont Dense models and Mixture-of-Experts (MoE) models.} \\
\midrule

\parbox[t]{\linewidth}{\linespread{1.2}\selectfont \textbf{Parallelism Strategy}} &
\parbox[t]{\linewidth}{\linespread{1.2}\selectfont \textbullet \hspace{0.2em} \textbf{Dense}: 3D parallelism (DP, PP, TP) with parallelism degrees denoted as $(d, p, t)$.
\newline
\textbullet \hspace{0.2em} \textbf{MoE}: A decoupled approach combining 2D (DP, PP) for Attention layers (denoted as $(d,p)$) and 3D (EDP, PP, EP) for MoE layers (denoted as $(d_e,p,e)$).} \\
\midrule

\parbox[t]{\linewidth}{\linespread{1.2}\selectfont \textbf{Parallelism \\ Configuration}} & 
\parbox[t]{\linewidth}{\linespread{1.2}\selectfont For a cluster of size $N$, configurations are enumerated as:
\newline
\textbullet \hspace{0.2em} \textbf{Dense}: All tuples $(d,p,t)$ where $d \cdot p \cdot t = N$.
\newline
\textbullet \hspace{0.2em} \textbf{MoE}: All tuples $(d_e,p,e)$ where $d_e \cdot p \cdot e = N$. The DP degree for Attention layers ($d$) is then determined by $d = d_e \cdot e$, which ensures that $d\cdot p = N$.} \\
\midrule

\parbox[t]{\linewidth}{\linespread{1.2}\selectfont \textbf{Parallelism Constraints}} & 
\parbox[t]{\linewidth}{\linespread{1.2}\selectfont \textbullet \hspace{0.2em} \textbf{PP degree ($p$)}: $2 \le p \le \frac{N}{16}$.
\newline
\textbullet \hspace{0.2em} \textbf{EP degree ($e$)}: $2 \le e \le \frac{N}{32}$ (with $\frac{e}{2}$ routed experts).
\newline
\textbullet \hspace{0.2em} \textbf{TP degree ($t$)}: $t \le$ GPUs per server (e.g., 8).} \\
\midrule

\parbox[t]{\linewidth}{\linespread{1.2}\selectfont \textbf{Workload Scaling}} & 
\parbox[t]{\linewidth}{\linespread{1.2}\selectfont Model parameters are scaled proportionally to parallelism degrees:
\newline
\textbullet \hspace{0.2em} \# Layers $\propto$ PP degree ($p$)
\newline
\textbullet \hspace{0.2em} Hidden Dimension $\propto$ TP degree ($t$)
\newline
\textbullet \hspace{0.2em} \# Experts $\propto$ EP degree ($e$)
\newline
\textbullet \hspace{0.2em} Batch Size $\propto$ DP degree of Attention layers ($d$)} \\
\midrule

\parbox[t]{\linewidth}{\linespread{1.2}\selectfont \textbf{Reference Models}} & 
\parbox[t]{\linewidth}{\linespread{1.2}\selectfont Scaling coefficients are derived from typical models: GPT-3 (for dense model) and DeepSeek-V3 (for MoE model).} \\
\bottomrule
\end{tabular}
\end{table}

\textbf{2. Traffic Modeling.} For each workload instance, its traffic demands are modeled as a sequence of communication primitives, where the parallel strategy dictates the primitive type and the model hyperparameter determines the traffic volume \cite{li_understanding_2024, jin_megascalemoe_2025}. Furthermore, similar to the approach in Megatron \cite{shoeybi_megatronlm_2020, narayanan_efficient_2021a, korthikanti_reducing_2023} and DeepSpeed \cite{rajbhandari_zero_2020}, data transfers for TP and DP are implemented as the Reduce-Scatter and All-Gather primitives using the Ring algorithm \cite{patarasuk_bandwidth_2009}. Data transfers for PP are implemented as Point-to-Point communication between GPUs in adjacent pipeline stages. Finally, the All-to-All communication for EP is modeled as a uniform traffic within each expert group, a simplification similar to the analysis in Rail-Only \cite{wang_railonly_2024}. To isolate the analysis on communication efficiency, GPU computation time is excluded from our study.

\textbf{3. Network Architectures.} Our analysis focuses on two architectural paradigms prevalent in hyper-scale AI data centers: the 3D-Torus and the Rail-Optimized architecture. The network configurations are specified in two parts: a baseline configuration for direct efficiency dissection (Section~\ref{sec:efficiency_dissection}), and a series of network parameter variation cases for investigating specific design trade-offs using our metrics (Section~\ref{sec:efficiency_analysis_parameters}).

\textit{\textbullet \hspace{0.2em} Baseline Network Configurations:} 

\textit{3D-Torus}. Each GPU is equipped with a 6-port switching chip. And assume the topology is reconfigurable via optical circuit switches (OCS), such that the communication within each parallel dimension in 3D parallelism is mapped to a dedicated physical dimension of the 3D-Torus.

\textit{Rail-Optimized}. An 8-GPU server configuration featuring a single-level intra-server switching fabric and an inter-server switching fabric (64-port switches), where the link bandwidth ratio (tiered bandwidth ratio) of the intra- to inter-server switching network is set to 9:1 (similar to the NVIDIA DGX H100 system \cite{ishii_nvlinknetwork_2022}). The cluster scale is set to 4096 for both architectures unless otherwise specified.

\textit{\textbullet \hspace{0.2em} Network Parameters:}

\begin{figure*}[!b]
    \centering
    \begin{subfigure}[c]{0.49\textwidth}
        \includegraphics[width=\linewidth]{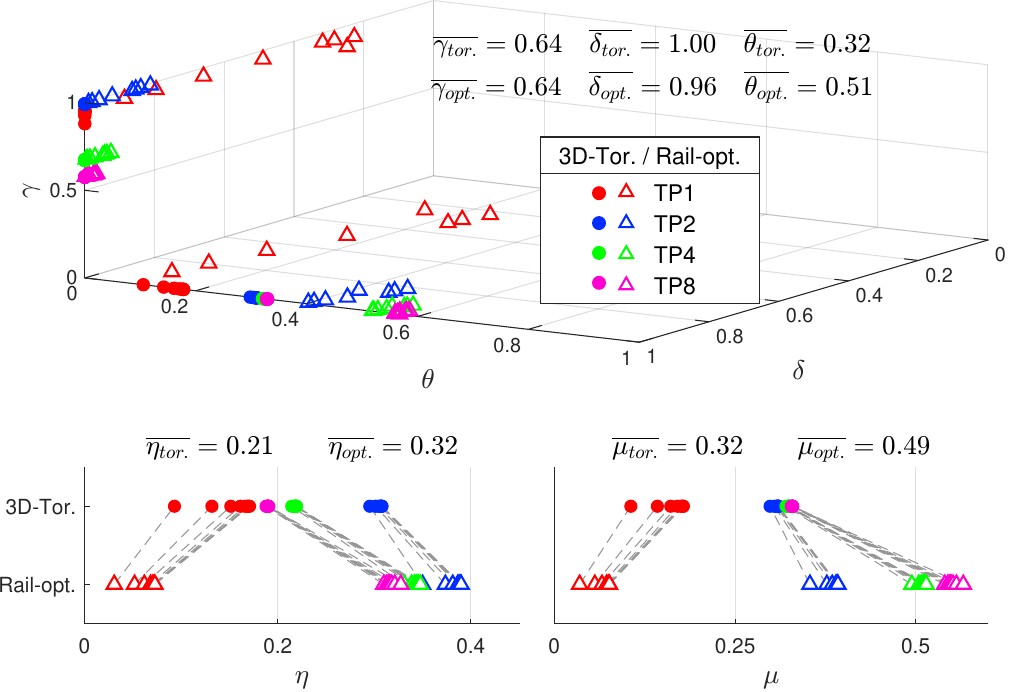}
        \caption{Dense model training workloads.}
        \label{dense_scatter}
    \end{subfigure}
    \begin{subfigure}[c]{0.49\textwidth}
        \includegraphics[width=\linewidth]{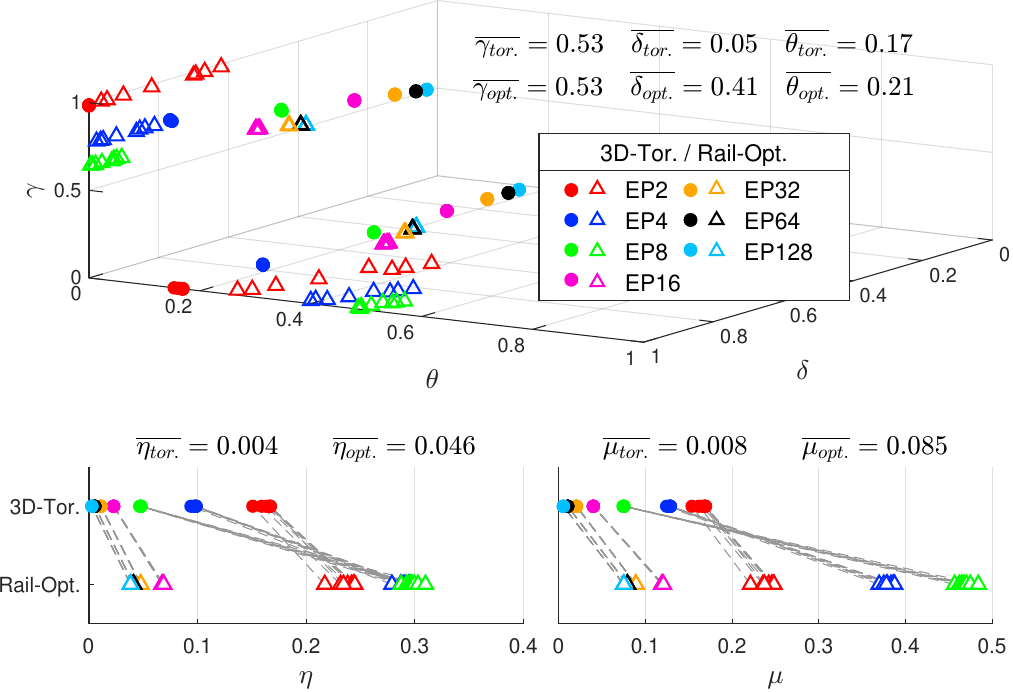}
        \caption{MoE model training workloads.}
        \label{moe_scatter}
    \end{subfigure}
    \caption{Efficiency dissection of 3D-Torus vs. Rail-Optimized for (a) dense and (b) MoE workloads on a 4096-GPU cluster. The top plot provides diagnosis of Data ($\gamma$), Routing ($\delta$), and Utilization ($\theta$) efficiency. The bottom plot presents the macroscopic assessment of Switching ($\eta$) and Network ($\mu$) Efficiency. Dashed lines link identical workloads across the two architectures.}
    \label{scatter}
    % \vspace{-10pt}
\end{figure*}

\textit{Case 1: Tiered Bandwidth Ratio.} Ratio ranging from 1:1 to 17:1 is evaluated for both architectures. For the 3D-Torus, we model this tiered ratio by increasing the port count along one dimension of the on-GPU switches, creating a high-bandwidth link that is dedicated to a traffic-intensive parallel dimension: EP for MoE models, or TP (or PP, when TP size is 1) for dense models.

\textit{Case 2: Server Size.} Server size ranging from 8 to 256-GPU is evaluated in Rail-Optimized architecture, with the assumption that a single-level intra-server switch fabric is maintained as the server size scales.

\textit{Case 3: In-Network Computing.} INC applied for the All-Reduce (substituting the All-Gather \& Reduce-Scatter primitives) within the TP phase by the intra-server switch of Rail-Optimized is evaluated. Our analysis focuses on this practical application, excluding two scenarios where practical implementation of INC is challenged: its application in 3D-Torus, where INC's tree-based communication pattern is incompatible with the distributed switch fabric \cite{ding_photonic_2025}, and its use for the All-to-All combine in MoE workloads, which is challenged by the fine-grained reduction scope \cite{zhao_insights_2025a}.

\textit{Case 4: Cluster Scale.} Cluster size scales from 512 to 65,536 GPUs is evaluated for both architectures. In particular, for the Rail-Optimized, multi-plane designs (single-, dual-, quad-, and octa-plane) analogous to the HPN architecture \cite{qian_alibaba_2024} are also included in this scalability analysis.  

\subsection{Efficiency Dissection of Baseline Network Architectures}
\label{sec:efficiency_dissection}

Fig.~\ref{scatter} presents the dissection of 3D-Torus and Rail-Optimized architectures on a 4096-GPU cluster, proceeding from a fine-grained diagnosis of data, routing, and resource utilization efficiency ($\gamma, \delta, \theta$) to a macroscopic assessment of overall performance ($\eta, \mu$) under diverse dense and MoE model workloads.

\textit{1) Observations from the fine-grained Data, Routing, and Port Utilization Efficiency:}

\textit{1a. Data Efficiency (${\gamma}$) reveals that redundant reception of data is a considerable bottleneck that constrains the end-to-end communication efficiency.} In particular, the redundant data reception $r_\text{byte}$ in implementations of Reduce-Scatter/All-Reduce and All-to-All combine without in-network computing, leads to lower $\gamma$ for workloads featuring more reduction-intensive communication (e.g., those with larger TP groups or larger EP groups). In aggregate, both the 3D-Torus and Rail-Optimized exhibit low $\bar{\gamma}$ of 0.64 for dense model workloads (Fig.~\ref{dense_scatter}) and 0.53 for MoE model workloads (Fig.~\ref{moe_scatter}). This data redundancy imposes a hard ceiling on end-to-end communication efficiency.

\textit{1b. Routing Efficiency (${\delta}$) highlights the challenge of MoE models for existing network architectures and underscores the deficiency of 3D-Torus in this scenario.} Specifically, multi-hop transmissions (i.e., large $\bar{n}_{\text{fwd}}$) would result in a small $\delta$, such as the cases of All-to-All traffic for MoE workloads in a 3D-Torus or traffic traversing a Leaf-Spine-Leaf path in a Rail-Optimized with an EP size $>$ 8. Overall, these cases cause $\bar{\delta}$ on the 3D-Torus to plummet from {1.00} for dense workloads --- where its reconfigurability provides perfectly matched single-hop paths for PTD parallelism --- to {0.05} for MoE workloads. For the Rail-Optimized, it decreases from {0.96} --- where dominant TP traffic is confined within a server --- to {0.41}. This sharp decline, particularly for the 3D-Torus, reveals their topological inadequacy for handling large-scale All-to-All traffic in MoE workloads.

\textit{1c. Port Utilization (${\theta}$) reveals the superiority of Rail-Optimized and highlights the challenges posed by MoE models in terms of resource utilization efficiency.} To be specific, the Rail-Optimized's tiered bandwidth design has an advantage over the 3D-Torus's symmetric design, as its high bandwidth intra-server subnetworks achieve higher $\theta_{\text{spatial}}$ when processing intensive localized traffic (e.g., TP traffic within a server). This leads to a superior $\bar{\theta}$ on dense model workloads ({0.51} vs. {0.32}). However, this advantage diminishes for MoE workloads, as cross-server All-to-All traffic underutilizes the fabric within the server, lowering $\theta_{\text{temporal}}$. In this scenario, both systems face a considerable challenge, with their $\bar{\theta}$ values dropping sharply to {0.21} and {0.17} respectively.

\textit{2) Synthesizing fine-grained factors by holistic efficiency metrics:}

\begin{figure*}[!b]
    \centering
    \begin{subfigure}[c]{0.4\textwidth}
        \includegraphics[width=\linewidth]{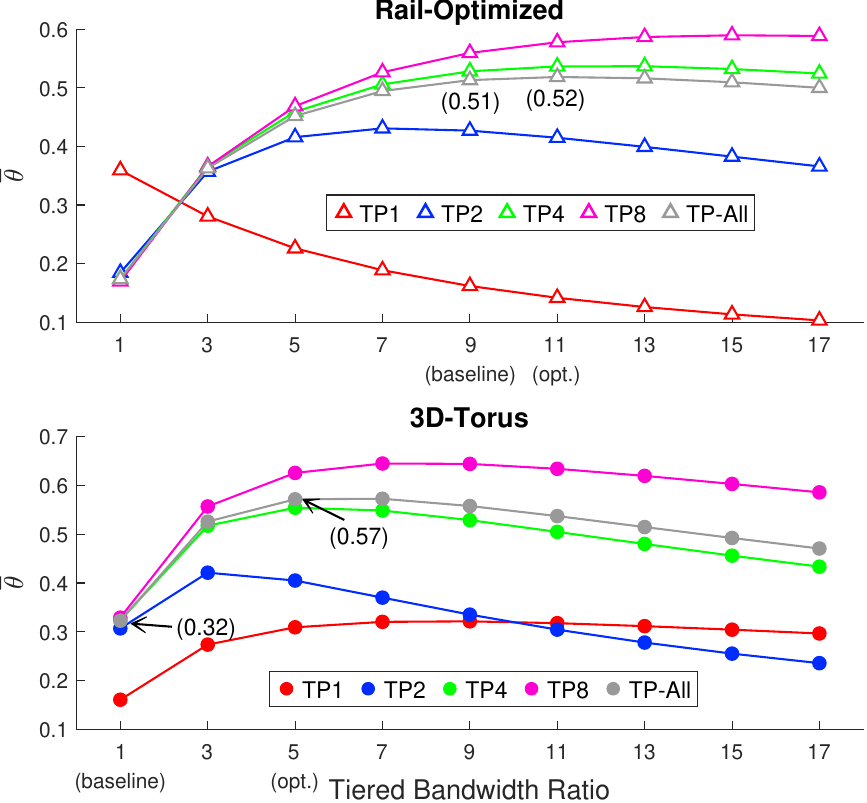}
        \caption{$\bar{\theta}$ vs. bandwidth ratio (dense models).}
        \label{BW_ratio_dense}
    \end{subfigure}
    \hspace{0.2cm}
    \begin{subfigure}[c]{0.4\textwidth}
        \includegraphics[width=\linewidth]{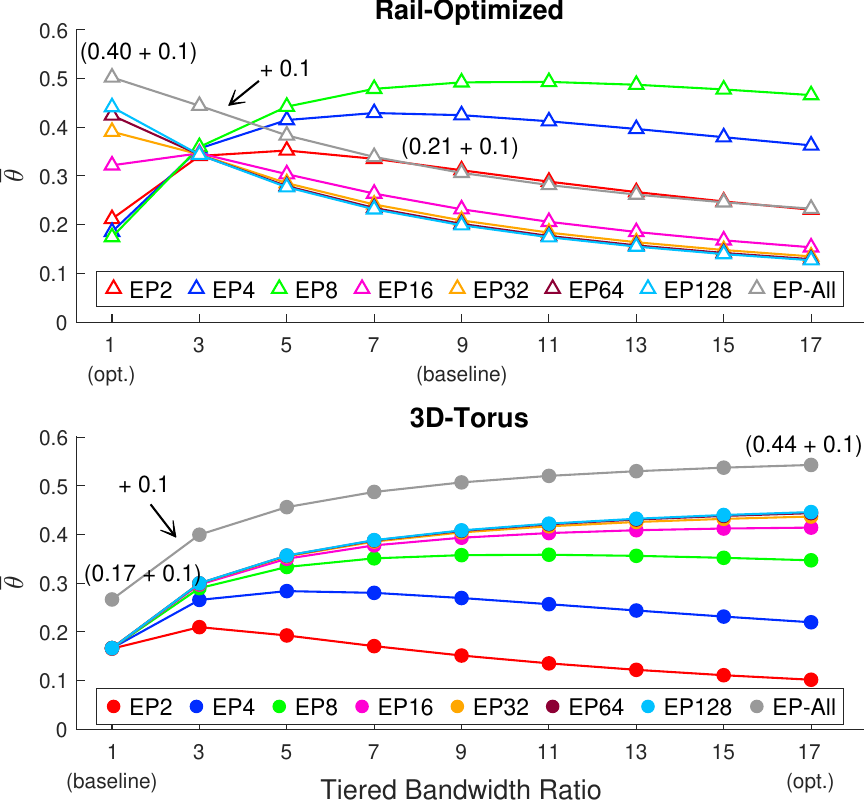}
        \caption{$\bar{\theta}$ vs. bandwidth ratio (MoE models).}
        \label{BW_ratio_moe}
    \end{subfigure}
    \caption{Impact of tiered bandwidth ratio on generalized Port Utilization ($\bar{\theta}$), evaluated across dense and MoE model workloads with different TP and EP sizes. For clarity, the EP-All curves are vertically shifted by +0.1.}
    \label{BW_ratio}
    % \vspace{-10pt}
\end{figure*}

The interplay of these fine-grained factors is synthesized by the holistic metrics ($\eta$ and $\mu$), leading to the following findings:

\textit{2a. The Rail-Optimized architecture exhibits higher efficiency.} Except for marginal cases at TP1, the Rail-Optimized design consistently achieves higher switching efficiency ($\eta$) across the remaining parallelism configurations. The generalized holistic metric $\bar{\eta}$ quantifies this overall advantage: Rail-Optimized outperforms 3D-Torus by {1.6x} in dense models ({0.32} vs. {0.21}), and dramatically extends this lead to {11.5x} in MoE models ({0.046} vs. {0.004}).

\textit{2b. MoE Workloads severely degrade efficiency in both architectures.} Both the network architectures suffer a sharp decline in switching efficiency ($\eta$) when shifting from dense model to MoE model workloads. Overall, $\bar{\eta}$ of the Rail-Optimized, drops by {85.6\%} from {0.32} to {0.046}. The degradation is more acute for the 3D-Torus, where $\bar{\eta}$ falls by {98.1\%}, from {0.21} to {0.004}.

\textit{2c. Data redundancy limits overall switching efficiency.} A persistent gap is observed between network efficiency ($\bar{\mu}$) and the overall switching efficiency ($\bar{\eta}$). Overall, exemplified by Rail-Optimized architecture, this efficiency gap causes a {35\%} efficiency loss for dense workloads (${\bar{\mu}=0.49}$ $\to$ ${\bar{\eta}=0.32}$) and a {46\%} loss for MoE workloads (${\bar{\mu}=0.085}$ $\to$ ${\bar{\eta}=0.046}$), underscoring a fundamental limitation that network-level optimizations alone cannot overcome.

\subsection{Efficiency Analysis of Varying Network Design Parameters}
\label{sec:efficiency_analysis_parameters}

The previous subsection applied the Switching Efficiency Framework for a top-down analysis of communication efficiency in fixed network architectures. This subsection extends its application to analyze how network parameters --- including tiered bandwidth ratio, server size, in-network computing, and cluster scale --- influence communication efficiency.

\textit{1) Impact of the Tiered Bandwidth Ratio on Switching Resource Utilization:}

\begin{figure*}[!h]
    \centering
    \begin{subfigure}[c]{0.32\textwidth}
        \includegraphics[width=\linewidth]{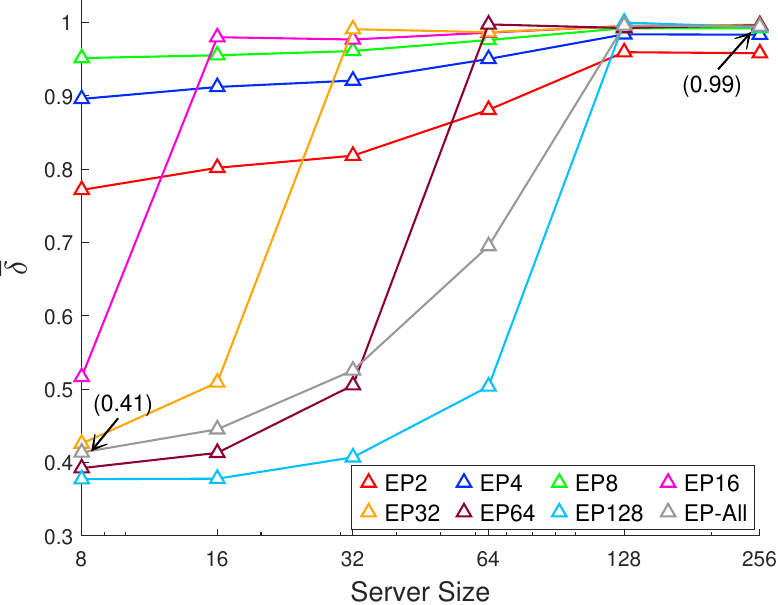}
        \caption{$\bar{\delta}$ vs. server size.}
        \label{server_size_delta}
    \end{subfigure}
    \begin{subfigure}[c]{0.32\textwidth}
        \includegraphics[width=\linewidth]{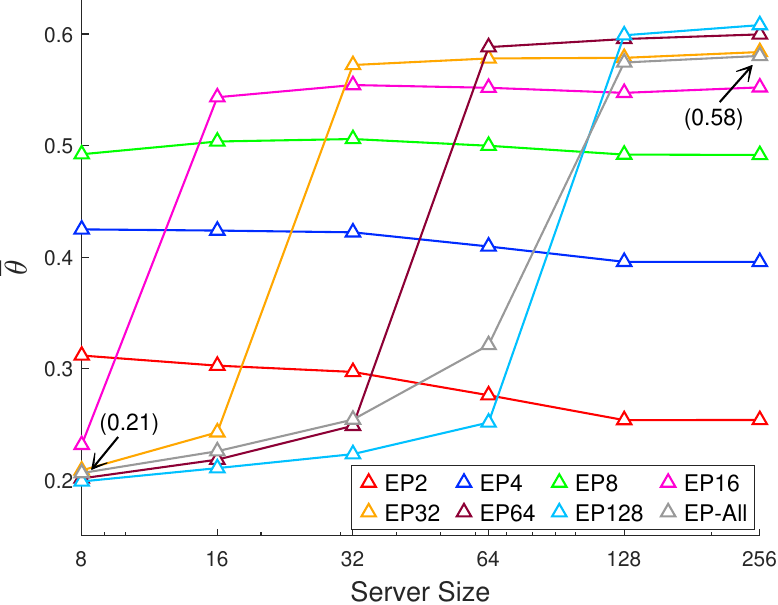}
        \caption{$\bar{\theta}$ vs. server size.}
        \label{server_size_theta}
    \end{subfigure}
    \begin{subfigure}[c]{0.32\textwidth}
        \includegraphics[width=\linewidth]{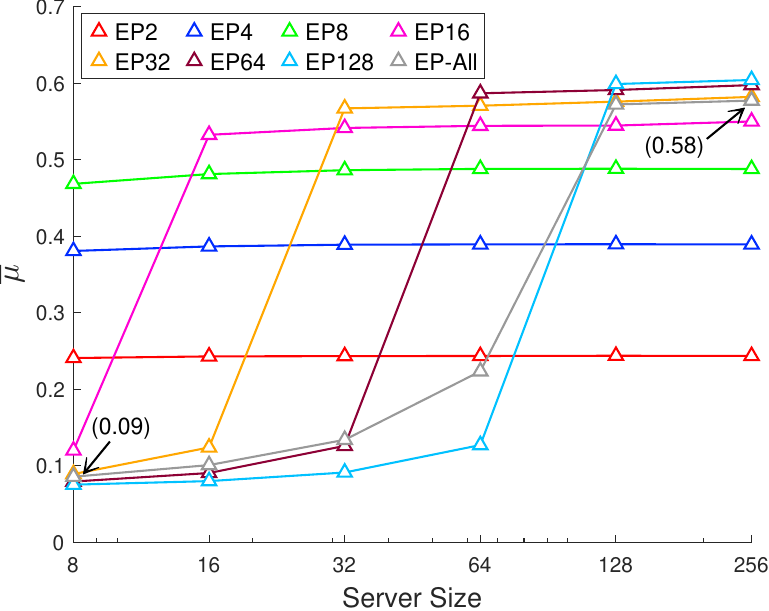}
        \caption{$\bar{\mu}$ vs. server size.}
        \label{server_size_mu}
    \end{subfigure}
    \caption{Influence of server size on communication efficiency under MoE training workload, with trends for generalized (a) Routing Efficiency ($\bar{\delta}$), (b) Port Utilization ($\bar{\theta}$), and (c) Network Efficiency ($\bar{\mu}$).}
    \label{server_size}
    % \vspace{-10pt}
\end{figure*}

To analyze the impact of the tiered bandwidth ratio on switching resource utilization, $\bar{\theta}$ is measured, revealing that strategically tuning the tiered bandwidth ratio leads to workload-dependent efficiency gains.

For dense workloads (Fig.~\ref{BW_ratio_dense}), the Rail-Optimized architecture experiences only a marginal gain from bandwidth ratio tuning, with its overall efficiency (TP-All) rising slightly from ${\bar{\theta} = 0.51}$ at baseline to {0.52} at optimum in our simulation configuration. In contrast, the 3D-Torus benefits far more significantly. At its optimal configuration, its overall port utilization (TP-All) increases by {78\%}, rising from a baseline of {0.32 to 0.57}. This substantial gain results from creating a high-bandwidth axis for the dominant parallel dimension (TP, or PP when TP=1), which enables the architecture to accommodate the traffic hierarchy.

For MoE workloads (Fig.~\ref{BW_ratio_moe}), the architectures show opposing responses to bandwidth hierarchy tuning. The 3D-Torus benefits significantly, with its overall port utilization (EP-All) increasing by {159\%} from a baseline $\bar{\theta}$ of {0.17 to 0.44} at the optimal configuration by dedicating a high-speed axis to EP communication. In contrast, for the Rail-Optimized architecture, a higher ratio is counterproductive. Its overall $\bar{\theta}$ (EP-All) drops from {0.40} at the 1:1 ratio to {0.21} at the baseline: when EP size exceeds server size, dominant all-to-all traffic is forced onto lower-bandwidth inter-server links, underutilizing the high-bandwidth intra-server fabric.

\textit{2) Influence of Server Size on Network-Level Efficiency for MoE model workloads:}

To assess the influence of server size on network-level communication efficiency for MoE workloads, whose dominant traffic (All-to-All traffic) regularly spans servers in practice, network-level metrics $\bar{\delta}$, $\bar{\theta}$, and $\bar{\mu}$ are leveraged to quantify this influence, showing that a larger server size in the Rail-Optimized leads to considerable efficiency improvements in the training of MoE models.

As shown in Fig.~\ref{server_size}, when the server size increases to match a specific EP size, the corresponding $\bar{\delta}$ and $\bar{\theta}$ for that workload become approximately {2 to 3 times} their baseline values, as the All-to-All communication in EP is now entirely contained within a single serve, which eliminates the multi-level switch forwarding and removes the utilization bottleneck formed by bandwidth disparity between intra- and inter-server.

In the general case, increasing the server size allows a greater portion of traffic to be resolved via single-level switches forwarding, rather than requiring multi-level forwarding, which promotes a rise in $\bar{\delta}$. But the reduced utilization of the top-level switches in turn lowers $\bar{\theta}$. $\bar{\mu}$ balances these two aspects, exhibiting a more stable trend. Overall (EP-All), as the server size increases from 8 to 256, {$\bar{\mu}$ rises sharply from 0.09 to 0.58}, demonstrating the substantial efficiency gains from enlarging the server size.

\textit{3) Efficiency Improvement from In-Network Computing  for Dense Model Workloads:}

To quantify the benefits of INC for dense model workloads, $\bar{\gamma}$ is measured to isolate its effect on eliminating redundant reception, and $\bar{\eta}$ is measured to analyze the overall impact, the results show that the application of INC eliminates redundant data reception, which translates to the increase in end-to-end communication efficiency.

\begin{figure}[h!]
    \centering
    \includegraphics[width=\columnwidth]{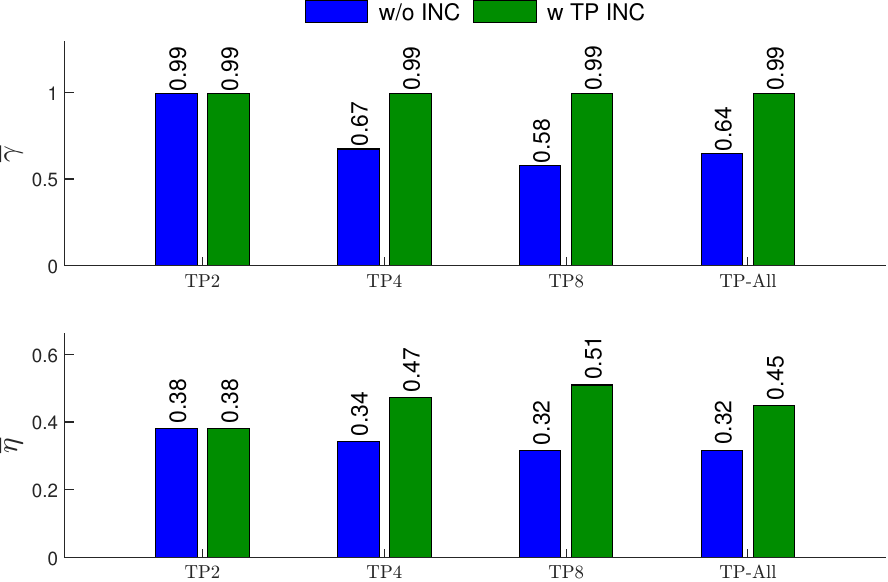} % Placeholder - use actual filename
    \caption{Efficiency improvement from the In-network computing under dense model training workloads in the Rail-Optimized architecture (note that the TP=1 case is omitted, as it involves no TP communication).}
    \label{INC}
    \vspace{0pt}
\end{figure}

As shown in Fig.~\ref{INC}, $\bar{\gamma}$ reaches a near-perfect {0.99} across all evaluated TP sizes (note that the Reduce-Scatter operation in DP communication still incurs redundant data reception). This high efficiency stems from the application of INC to the dominant TP communication, which eliminates redundant data reception inherent in the reduction process, thereby raising its data efficiency to the ideal value of 1. The improvement in data efficiency, in turn, boosts the overall $\bar{\eta}$ (TP-All) from {0.32 to 0.45}, demonstrating the gains in end-to-end communication efficiency from applying INC.

\textit{4) Impact of Network Architecture Scale on Overall Communication Efficiency}:

\begin{figure*}[!h]
    \centering
    \begin{subfigure}[c]{0.38\textwidth}
        \includegraphics[width=\linewidth]{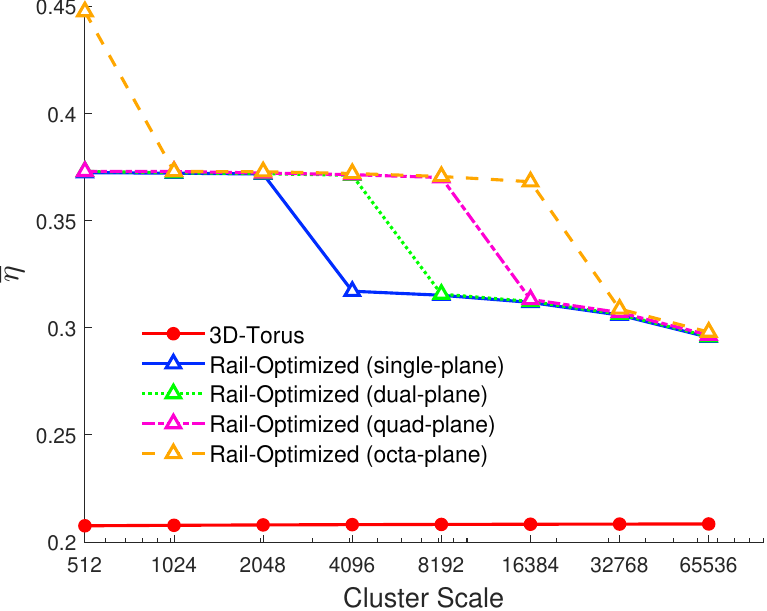}
        \caption{$\bar{\eta}$ vs. cluster scale (dense models).}
        \label{dense_scalability}
    \end{subfigure}
    \hspace{0.2cm}
    \begin{subfigure}[c]{0.38\textwidth}
        \includegraphics[width=\linewidth]{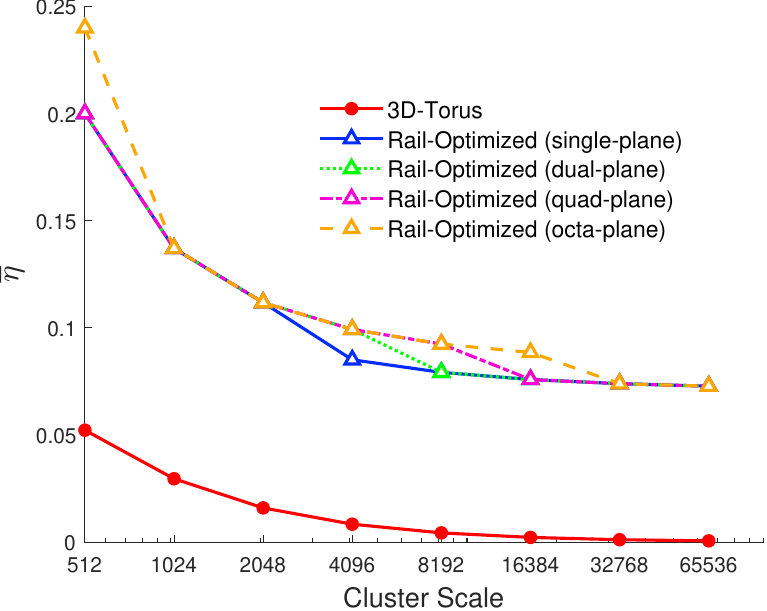}
        \caption{$\bar{\eta}$ vs. cluster scale (MoE models).}
        \label{moe_scalability}
    \end{subfigure}
    \caption{Generalized Switching Efficiency ($\bar{\eta}$) of 3D-Torus and Rail-Optimized architectures as cluster scales from 512 to 65,536 GPUs under (a) dense and (b) MoE model training workloads.}
    \label{scale}
    % \vspace{-10pt}
\end{figure*}

To investigate architectural scalability, $\bar{\eta}$ is employed to track overall communication efficiency across a wide range of cluster sizes, revealing that different architectures exhibit distinct efficiency degradation trends as the cluster size grows.

Under dense model workloads (Fig.~\ref{dense_scalability}), 3D-Torus maintains a persistently low $\bar{\eta}$ of {0.21}. This is because, while 3D-Torus provides a single-hop topology for the PTD-parallelism traffic --- a property that remains constant as it scales --- its symmetric fabric imposes a uniform bandwidth allocation, which is ill-suited to the imbalanced LLM training traffic. In contrast, while Rail-Optimized architectures exhibit a stepwise decline in efficiency, multi-plane designs delay this degradation. For instance, $\bar{\eta}$ of the single-plane architecture drops from 0.37 at 2048 GPUs to 0.32 at 4096 GPUs, reflecting the addition of a new switching layer that incurs inefficiency in both $\bar{\delta}$ and $\bar{\theta}$ (due to potential increased forwarding hops or idleness of upper-layer switches). The octa-plane design, however, sustains this high efficiency of ${\bar{\eta}\approx 0.37}$ up to 16384 GPUs, as multi-plane design delays the need to add switching layers to support cluster expansion.

Under the MoE model workloads (Fig.~\ref{moe_scalability}), all architectures exhibit a continuous decline in $\bar{\eta}$ when EP sizes scale with cluster size. For the 3D-Torus, $\bar{\eta}$ declines rapidly from {0.033} to below {0.001} for scaling from 512 to 16384 GPUs, because larger EP groups amplify the routing inefficiency of the All-to-All traffic on its neighbor-only connectivity topology. For the Rail-Optimized architectures, $\bar{\eta}$ also declines continuously, as growing EP groups force more All-to-All traffic onto inefficient, multi-hop inter-server links. Despite this overall decline, multi-plane designs prove more scalable, with the octa-plane design, for instance, maintaining a slight advantage over the single-plane as the cluster scales from 4096 to 16384 GPUs ($\bar{\eta}$ ranging from {0.053 to 0.046 vs. 0.046 to 0.039}). 

\section{Insights for Future Network Evolution}

Fig.~\ref{scatter} reveals that communication efficiency, across both holistic and fine-grained metrics, is highly dependent on communication-intensive strategies like TP and EP. Given that the communication arising from these strategies is typically mapped to the High-Bandwidth Domain (HBD, e.g., high bandwidth NVLink within the server), and that these HBDs constitute the majority of the network switching resources, we argue that co-optimizing the design and usage of HBDs is a pivotal pathway to improving communication efficiency.

\textit{Empowering the server-scale HBD in Rail-Optimized.} For the Rail-Optimized architecture, expanding the server size to encompass the entire All-to-All communication within an EP group effectively boosts both port utilization ($\theta$) and routing efficiency ($\delta$) for MoE model training, as shown in Fig.~\ref{server_size}. Similarly, equipping intra-server switches with INC capabilities prevents GPUs from receiving redundant, unreduced data during TP, leading to a significant improvement in data efficiency ($\gamma$), as shown in Fig.~\ref{INC}. Both the server-level enhancements yield more profound efficiency returns compared to adjusting the bandwidth resource allocation (Fig.~\ref{BW_ratio}) or adopting a multi-plane switching plane design (Fig.~\ref{scale}).

\textit{Curating workloads for the neighbor-only connectivity HBD of 3D-Torus.} The 3D-Torus equips each accelerator with low-radix switches to form a distributed switching topology \cite{zu_resiliency_2024}, forming an HBD that spans the entire cluster but offers limited connectivity flexibility. However, executing workloads that incorporate 3D hybrid parallelism on a 3D-Torus requires a one-to-one mapping between the orthogonal parallel dimensions and the network's physical dimensions \cite{jouppi_tpu_2023}, which leads to inefficiency due to bandwidth fragmentation, as quantified by $\theta$. Furthermore, mapping the All-to-All traffic of EP to a single dimension of the Torus necessitates multi-hop transmissions, resulting in routing inefficiency, as quantified by $\delta$. In the absence of switch chips that permit flexible bandwidth allocation across dimensions (an alternative explored in Fig.~\ref{BW_ratio}), specializing workloads becomes a necessary strategy to enhance efficiency. For instance, dedicating a single Torus network exclusively to model parallelism would allow its high-volume data transfers to leverage switch ports across more network dimensions and reduce the communication diameter, a practice exemplified by Google in training its Gemini models \cite{rohananil_gemini_2025, comanici_gemini_2025}.

\section{Conclusion and Future Work}

This paper introduces the Switching Efficiency Framework, a hierarchical metrics set for top-down analysis of how effectively network switching capacity is converted into computationally useful data throughput in LLM training. By shifting the evaluation paradigm from monolithic performance benchmarking to diagnostic analysis, this work provides actionable guidance for the resource-efficient design of future AIDCs. Applying this framework, we revealed the (mis)alignment between the design patterns of mainstream architectures such as 3D-Torus and Rail-Optimized, and the structured traffic patterns they support. Moreover, the fine-grained metrics enabled us to diagnose the root causes of their specific efficiency bottlenecks and identify potential optimization pathways, thereby showcasing the practical utility of the proposed framework.

%-------------------------------------------------- Acknowledgements Section -------------------------------------------------------%

\section{Acknowledgements}
This work was supported in part by the National Key Research and Development Project of China under Grant 2024YFB2908301 and in part by the National Natural Science Foundation of China (NSFC) under Grant 62331017. (Corresponding author: Weiqiang Sun.)

\footnotesize
\printbibliography
\end{document}